\documentclass[preprint,authoryear,12pt]{elsarticle}
\usepackage{amsmath,amssymb,amsbsy}
\journal{Icarus}

\begin{document}

\begin{frontmatter}
\title{Ejecta Transfer in the Pluto System}
\author{Simon B. Porter\corref{cor1}}
\ead{porter@boulder.swri.edu}
\author{William M. Grundy}
\address{Lowell Observatory, 1400 W Mars Hill Rd, Flagstaff, AZ 86001}
\cortext[cor1]{Now at Southwest Research Institute}

\begin{abstract}
   The small satellites of the Pluto system (Styx, Nix, Kerberos, and Hydra) have very low surface escape velocities,
   and impacts should therefore eject a large amount of material from their surfaces.
   We show that most of this material then escapes from the Pluto system, though a significant fraction
   collects on the surfaces of Pluto and Charon.
   The velocity at which the dust is ejected from the surfaces of the small satellites strongly determines 
   which object it is likely to hit, and where on the surfaces of Pluto and Charon it is most likely to impact.
   We also show that the presence of an atmosphere around Pluto eliminates most particle size effects and increases 
   the number of dust impacts on Pluto. 
   In total, Pluto and Charon may have accumulated several centimeters of small-satellite dust on their surfaces,
   which could be observed by the New Horizons spacecraft.
\end{abstract}

\end{frontmatter}

\pagebreak

\section{Introduction}
Pluto and its satellites form a uniquely complex dynamical system.
Pluto and Charon are a true binary system with a mass ratio of 8.6 ($\pm$0.5) to 1,
and a center of mass 890 km (0.77 Pluto radii) above the surface of Pluto \citep{Tholen2008}.
Around them orbit four known small satellites, Nix and Hydra \citep{Weaver2006}, Kerberos \citep[formerly P4,][]{Showalter2011}, 
and Styx \citep[formerly P5,][]{Showalter2012}.
The small satellites follow near-circular orbits centered on the Pluto-Charon center of mass and coplanar with
the Pluto-Charon orbital plane.
These orbits are stable, though only a slight change in eccentricity or inclination would lead to chaotic
trajectories \citep{Youdin2012}.
It is therefore not trivial to predict the behavior of dust in the Pluto system.

Impacts onto Jupiter's innermost four satellites (from inner to outer, Metis, Adrastea, Amalthea, and Thebe)
produce faint rings of short-lived dust particles \citep{Burns1999}.
This dust is typically ejecta from impacts of interplanetary dust particles (IDPs) onto the inner satellites,
and has a mean grain size of 5 $\mu{\rm m}$.
Observations from the \textit{Spitzer} space telescope have shown that Saturn hosts a similar (but much larger)
impact generated ring, formed from material ejected from the retrograde irregular satellite Phoebe \citep{Verbiscer2009}.
The Phoebe ring extends inwards from Phoebe's orbit to at least 128 Saturn radii, and likely as far as 
the outermost regular satellite of Saturn, Iapetus, at 60 Saturn radii \citep{Verbiscer2009}.
The ring material is swept up by Iapetus, 
accumulating over the age of the solar system approximately 20 cm of material,
globally averaged \citep{Tamayo2011}.
Importantly, though, the Phoebe ring impacts are not isotropic on the surface of Iapetus, but 
strongly favor the leading hemisphere.
Thermal segregation caused by this accumulation of dark Phoebe material on the leading hemisphere contrasting with
the relatively clean water ice trailing hemisphere explains the wildly different albedo of the two hemispheres
\citep{Spencer2010}.

The transfer of dust between satellites has also been used to explain the surfaces of other icy satellites.
\citet{Bottke2013} suggested that the dark albedos of the Galilean satellites are a result of
dust ejected from the small irregular satellites of Jupiter.
\citet{Schenk2011} trace the outer edges of the E-ring generated by Enceladus's plumes and find that 
E-ring dust could be responsible for equatorial features seen on Rhea.
\cite{Tamayo2013} showed that the dust influx from irregular satellites may cause the leading/trailing
color asymmetry seen on Uranus's irregular satellites.
All icy satellites should therefore be considered in the context of their dust environment.

\citet{Thiessenhusen2002} suggested that Pluto and Charon are surrounded by a cloud of dust ejected from
their surfaces, similar to Jupiter's impact-generated rings.
This was difficult to show for dust from Pluto or Charon, though, as it would require a significant fraction of its impact
velocity to escape from their surfaces.
\citet{Stern2006} then suggested that impact ejecta dust from the newly discovered small satellites
Nix and Hydra could produce temporary dust rings.
The small satellites have much lower surface escape velocities, and so could potentially eject much more dust into the system.
\citet{Steffl2007} measured an upper limit for any dust in the system which they used to
limit the median dust particle lifetime to approximately 900 years.
\citet{Poppe2011} and \citet{dosSantos2013} both performed numerical simulations of dust particles ejected from
Nix and Hydra and perturbed by of solar radiation pressure.
\citet{dosSantos2013} showed that only a small fraction of dust particles smaller than 10 $\mu{\rm m}$ survive for more than 100 years,
while \citet{Poppe2011} found much longer lifetimes for particles larger than 10 $\mu{\rm m}$.

\citet{Stern2009} suggested that this dust may also be swept up by the other objects in the system.
\citet{Poppe2011} and \citet{dosSantos2013} then showed that the small satellite ejecta generally transfers inward, producing secondary impacts on Pluto and Charon.
Here, we reproduce their simulations, but focusing on the impacts rather than the long-lived dust trajectories.
In addition to estimating the fraction of small satellite ejecta which impacts Pluto and Charon,
we also estimate the spatial distribution of those impacts on Pluto and Charon.
Over time, the small satellites could have transferred a considerable amount of material to the surfaces of Pluto and Charon.
We show that spatial distribution of the those impacts is sufficiently unique that it might observable by NASA's 
\textit{New Horizons} spacecraft when it flies past Pluto and Charon.
We also include for the first time the effect of air drag from Pluto's atmosphere, and show that it reduces
the number of long-period dust particles interior to Charon's orbit.
And we directly compare these ejecta dust simulations to the trajectories of interplanetary 
dust particles through the system.

\section{Simulation Methods}
We performed a large number of numerical simulations of dust trajectories within the Pluto system, 
both for interplanetary dust particles (IDPs) and for those ejected from Nix and Hydra.
In these simulations, we randomized the start time and the initial direction of the dust particle, and then followed the particle
until it either left the Pluto system or impacted another object.
The IDPs started on hyperbolic orbits through the system typical of dust migrating inwards from the Kuiper Belt, 
while the ejecta particles were launched from the surfaces of their source satellites.

In the simulations, we included three forces: gravity, solar radiation pressure, and aerodynamic drag from Pluto's atmosphere.
Gravity from the Sun was included, as was mutual gravitation between Pluto, its satellites, and the dust particle.
The Sun was the coordinate origin of the simulations, and the Pluto system was placed in orbit around the Sun.
While solar gravity is relatively weak for dust close to the Pluto-Charon barycenter, it can be influential on dust which is excited to 
highly eccentric orbits after a close encounter with Pluto.
Solar gravity is one tenth as strong as Pluto-Charon gravity at 134 Pluto radii from the barycenter, or 2.4 times the size of Hydra's orbit.
A dust particle would therefore not need to be excited much beyond its source satellite in order to experience significant solar gravity perturbations.

For the radiation pressure, we assumed the dust particle was roughly spherical, and so the acceleration on the particle was \citep{Burns1979}:
\begin{align}
   \label{radpress}
   \ddot{\vec{r}}_{rp} &= \frac{3}{8\pi} \frac{L_{\odot} Q_{pr}}{c \rho} \frac{\vec{r_{\odot}}}{r_{\odot}^3 D} \\
   \label{radpress_a}
   \ddot{\vec{r}}_{rp} &\approx 1.1\times10^{12} \frac{km^4}{day^2}~\frac{\vec{r_{\odot}}}{r_{\odot}^3 D}
\end{align}
where $L_{\odot}$ is the solar luminosity, $Q_{pr}$ is the radiation pressure efficiency (assumed to be unity),
$c$ is the speed of light in a vacuum, $\rho$ is the density of the dust grain (assumed to be 1 g/cm$^3$),
$\vec{r_{\odot}}$ is a vector from the Sun to the dust, and $D$ is the diameter of the dust grain in $\mu$m.
\citet{Burns1979} showed that $Q_{pr}$ is close to one for particle sizes from 0.5-10 $\mu$m.
Larger particles might have internal transmission \citep{Hapke1993}, which increases forward scattering and thus lowers $Q_{pr}$.
Equation \ref{radpress_a} therefore provides close to an upper limit of the radiation pressure force.
At Pluto's distance from the Sun and with $Q_{pr}$ assumed to be unity,
solar radiation pressure and solar gravity almost perfectly cancel each other out for 1 $\mu$m particles 
(since the two forces are near-equal magnitude and precisely opposite directions).
For larger particles, solar radiation pressure is weaker due to the higher amount of mass per cross-sectional area,
and solar gravity dominates over radiation pressure.
Following after \citet{dosSantos2013}, we did not include Poynting-Robertson forces nor any non-spherical gravity
fields, as both are very small for this case.

Pluto's atmosphere is thin, but its drag on small particles can be significant.
From stellar occultations, the atmosphere is nearly isothermal, and so the density profile 
can be approximately modeled as exponential with altitude \citep{Elliot1992}.
We assumed a scale height of 55 km and a density of 7${\times}10^{-9}$ g/cm$^3$ at 1275 km from Pluto's center of figure \citep{Young2008a}.
This formulation reaches double-float (64-bit) machine precision at approximately the orbit of Charon,
effectively truncating the atmosphere at this distance.
However, since the density falls off exponentially, a dust particle must be within a few thousand kilometers of Pluto for drag to be significant.
We ran most of the simulations both with and without a Pluto atmosphere, in order to isolate the effects of drag on the dust particles.

Since the dust particles approach Pluto at high speeds (several hundred m/s),
we used Rayleigh drag (proportional to the square of the velocity) to approximate the aerodynamic forces on the dust particles.
In addition, we again assumed the dust particles were roughly spherical for the purposes of estimating their cross-section.
The acceleration due to drag on the particle was then:
\begin{align}
   \label{drag}
   \ddot{\vec{r}}_{drag} &= -\frac{3}{4}\frac{\rho_{atm}(\vec{r_{P}})}{\rho_{dust}}\frac{C_D}{D}v_{P}^2\hat{v_{P}}
\end{align}
where $\rho_{atm}(\vec{r_{P}})$ is the density of Pluto's atmosphere as a function of distance from Pluto,
$\rho_{dust}$ is the density of dust particle, $C_D$ is the coefficient of drag for the dust particle,
and $\vec{v_P}$ is the velocity of the particle relative to Pluto.
We assumed a mean density for the particle of 1.0 g/cm$^3$, appropriate for mixture of water ice, organics, and silicates 
with porosity and the rough shapes seen in IDPs collected in the Earth's atmosphere \citep{Flynn1994}.
Likewise, because of this rough shape, we assumed a coefficient of drag of 1.0.
With these physical assumptions and drag from Equation \ref{drag}, the terminal velocity of the dust particles at the surface of Pluto
can be approximated as  $v_{terminal} \approx \sqrt{14 D}$ m/s, where $D$ is the dust diameter in $\mu$m.
Thus, the terminal velocity of a 1 $\mu$m particle is 4 m/s, while the terminal velocity of a 1 mm particle is 120 m/s.

Gravity and radiation pressure are conservative forces, but drag is not.
Thus, we could not use a traditional symplectic integrator, and instead we employed the integration scheme of \citet{Cash1990}.
This is a fifth-order Runge-Kutta method with several embedded lower orders. 
These lower orders allow for the calculation of three different error terms for one integration step, 
and can thus detect sudden changes in otherwise smooth functions.
Such a sudden change occurred in our simulations when a dust particle in a barycentric orbit had a close encounter with Pluto or Charon.
The integrator was designed to be precise enough during 
these close approaches to accurately capture the impacts and their locations.
We performed the simulations in units of kilometers for distance and kilometers/day for speed in order to 
keep the magnitude of the forces large and therefore
minimize the amount of rounding errors per integration step.

We used the state vectors and masses for Pluto, Charon, Nix, and Hydra from \citet{Tholen2008} as initial conditions
and evolved them forward or backward using the N-Body integrator for a random amount of time, up to $10^6$ days.
We assumed that both Nix and Hydra have diameters of 100 km, which is within the uncertainty from \citet{Tholen2008}.
We then chose a random unit vector for the dust ejection direction ($\vec{u}_{rand}$), using the algorithm given in \citet{Knop1970}.
The IDPs were injected into the system from a random point on the Hill sphere,
with initial velocity vectors pointed randomly within 10 degrees of the center of mass,
causing them to within about $2{\times}10^6$ km of the barycenter (or about 25 times Hydra's semimajor axis).

The small satellite ejecta particles were ejected vertically from random locations on the surfaces of their source satellites.
The ejecta particle's initial position ($\vec{r}_{dust}$) and velocity ($\vec{v}_{dust}$) were then:
\begin{align}
   \label{rdust}
   \vec{r}_{dust} &= \vec{r}_{sat} + ((R_{sat}+100 m) \cdot \hat{u}_{rand})\\
   \label{vdust}
   \vec{v}_{dust} &= \vec{v}_{sat} + (v_{ejection} \cdot \hat{u}_{rand})
\end{align}
where $\hat{u}_{rand}$ is a random unit vector.
This starts the dust particle 100 meters above the surface of the source satellite 
(to provide a buffer against an impact detection on a poorly-adapted initial timestep), 
and travelling vertically up from the local surface (assuming the source is spherical).
The ejection speed ($v_{ejection}$) and dust diameter are given values, and remain constant for each set of simulations.
We used particle sizes of 1, 10, 100, and 1000 $\mu$m.

We varied the ejection velocity of the dust from the small satellite between the satellite's 
surface escape velocity ($\approx$30 m/s) and the velocity where all the dust is ejected faster then the 
Pluto system escape velocity ($\approx$400 m/s).
Since the small satellites of Pluto all have orbital velocities slower than 200 m/s, 
for most of the velocity range, the dust was ejected faster than the source satellite's orbital velocity.
All but the slowest particles were ejected with a significantly different velocity vector to their parent,
placing them on more eccentric and inclined orbits.

Impacts, when the dust particle is within the nominal radius of another object, are checked for at the start of each timestep.
If an impact is detected, the simulation is stopped and the final state output.
All the impacts were detected within a few km of the nominal radius of the object, and most within a few meters.
Thus, the integrator was sufficiently responsive to detect impacts at or near the surfaces.
The simulations were also stopped if the dust exceeded a maximum distance from Pluto (1.2$\times$10$^7$ km, approximately twice Pluto's Hill Radius),
or if the dust exceeded the maximum allowed lifetime (10$^6$ days).

\section{Simulation Results}

There are three basic fates for a dust particle flying through the Pluto system: 
ejection from the Pluto system,
reimpacting their original source satellite (for the ejecta particles),
or impacting the surfaces of either Pluto or Charon.
All of the dust ejected from the small satellites faster than 400 m/s escaped from the system,
as did most of the dust ejected at slower velocities.
In addition to the dust particles that are initially ejected fast enough to escape,
many slower ejecta particles also escaped after having a close encounter with either Pluto or Charon.
On the other hand, nearly all of the dust ejected from the small satellites slower than 40 m/s reimpacted
onto its source satellite. 
This speed is faster than the escape velocity from the small satellites, 
but is not fast enough to avoid being coorbital with the source, and reimpacting after a few orbits.
Our simulations did not produce any of the ``sailboat orbits'' proposed by \citet{GiuliattiWinter2010},
nor any dust orbits with an apoapse close enough to Pluto that could have evolved into that region.

The third potential fate is for the dust particle to impact the surface of another body, chiefly Pluto or Charon.
The probability of impact was mostly a function of ejection velocity, as can be seen in Figures \ref{fig:velocity_nodrag} and \ref{fig:velocity}.
For a given ejection velocity and particle size, this only occurred in less than 13\% of
ejection simulations, and typically much less.
Charon impacts were primarily low-speed ejecta just fast enough to prevent reimpacting the source.
Pluto-impacting dust, on the other hand, was typically ejected at slightly higher velocities.
There was some exchange of dust between the small satellites for ejecta speeds slower than 50 m/s,
but this occured in less than 0.1\% of simulations with ejecta faster than 50 m/s.

The impact velocities on Charon were slightly faster than Charon's surface escape velocity (575 m/s).
However, since the surface escape velocity of Pluto (1229 m/s) is much faster than the terminal velocity of the 
dust particles, all the Pluto impacts were at that particle's terminal velocity (4 to 120 m/s).
Thus, the dust impacts on Charon are much more energetic than Pluto.
This combined with Charon's smaller surface area (one third of Pluto) to mean that Charon may receive much more micrometeorite space weathering than Pluto.

In the simulations of IDPs, approximately 3\% impacted Pluto and 0.4\% impacted Charon, independent of particle size.
Pluto's atmosphere and larger gravitational focusing thus gave it an effective cross-section 7.5 times larger than
Charon (the physical cross-section ratio is approximately 3.6:1).
Accounting for the difference in surface area, Pluto should then receive approximately twice the amount of IDPs per unit area as Charon.
However, the impacts on Charon are not slowed by atmospheric drag, and the IDP impact kinetic energy flux is higher on Charon than Pluto.
The IDP impacts on Pluto were more common on the anti-Charon hemisphere, but did occur over the entire surface.

\subsection{Ejection Velocity Effects}

All the ejected dust particles follow orbits dominated by the Pluto-Charon barycenter.
If the particle's ejection velocity is fast enough, it will leave the system on a hyperbolic trajectory.
Since the small satellites are on almost-circular orbits relative to the barycenter,
the minimum escape velocity is $\sqrt{2}$ times their orbital velocity.
The orbital velocity of Nix is about 141 m/s and 123 m/s for Hydra, and so the minimum escape velocities are
about 200 m/s for Nix and 173 m/s for Hydra.
The smallest-magnitude ejection velocity vector for direct escape is from the center of the leading hemisphere
which equates to an ejection speed of 59 m/s for Nix and 50 m/s for Hydra.
Any dust ejected slower than this can only escape through a close encounter with Pluto or Charon.
Dust ejected from the trailing hemisphere of the small satellites, on the other hand,
must be going fast enough to both cancel out the satellite's velocity and then escape from the system.
This requires an ejection speed of 340 m/s for Nix and 296 m/s for Hydra.
Any dust ejected faster than this, regardless of direction, will initially be on a hyperbolic escape trajectory from the Pluto system.

Ignoring three-body effects, the minimum energy trajectory from the small satellites to Pluto 
requires a delta v of about 95 m/s, and 40 m/s to Charon.
However, Figure \ref{fig:velocity} shows that a small fraction of particles which were ejected slower than these velocities still impacted Pluto or Charon.
The slow (ejection velocity \textless40 m/s) 
Charon-impacting particles stayed on bound trajectories around the barycenter for several orbits before being swept up.
These particles initially had periapses exterior to Charon's, but close enough that Charon 
either perturbed them into a Charon-crossing orbit or directly focused them on its surface.

For Pluto impactors, there was an even greater enhancement below the minimum energy transfer (95 m/s) because of Charon scattering events.
The Pluto impacts show a low-velocity secondary peak corresponding to the peak in Charon impacts.
These are dust particles which approach Charon but do not impact, and are instead scattered inwards on a Pluto-crossing trajectory
(much like a spacecraft gravity assist).
Thus, even very low velocity ejecta from the small satellites can impact Pluto.

As Figure \ref{fig:lifetimes} shows, 
the median lifetime of the ejecta particles as a whole decreased with higher ejection velocities, greater than 95 m/s.
Since a particle ejected at the minimum escape velocity from Nix or Hydra would take
7330 days to reach the arbitrary maximum distance of 1.2$\times$10$^7$ km,
the apparent lifetimes of the escaping ejecta particles were all at least thousands of days.
The dust particles that impacted either Pluto or Charon (the dotted lines in Figure \ref{fig:lifetimes})
typically lasted a few orbits around the barycenter before being swept up.
The median lifetime of the impactors drops off as the ejection velocity approaches the point where all 
dust is ejected faster than escape velocity, implying that most impact of these particles 
Pluto on their first (and only) periapse passage.

\subsection{Radiation Pressure and Atmospheric Drag Effects}

We included three addition forces on the particles in addition to gravity from Pluto and its satellites.
All the simulations included gravity from the Sun, as well as solar radiation appropriate to the size of the particle.
These forces were small, and only influential on the small fraction of eccentric orbits with apoapses close to the edge of Pluto's Hill sphere.
The addition of gas drag from Pluto's atmosphere slows down particles which pass interior of Charon's orbit,
increasing the number of Pluto impacts and preventing dust from entering into orbits eccentric enough to feel solar perturbations.

To probe the effect of radiation pressure on the dust trajectories,
we repeated all of our simulations for particle grain sizes of 1, 10, 100, and 1000 ${\rm\mu{m}}$.
From Equation \ref{radpress_a}, this approximately corresponds to radiation pressure accelerations of
4.5$\times10^{-6}$ m/s$^2$, 4.5$\times10^{-7}$ m/s$^2$, 4.5$\times10^{-8}$ m/s$^2$, and 4.5$\times10^{-9}$ m/s$^2$ respectively.
For comparison, acceleration due to solar gravity at Pluto is 4.1$\times10^{-6}$ m/s$^2$, and so the two solar forces are
comparable for the 1 $\mu$m particles, and dominated by solar gravity for larger particles.

\begin{figure}[p!]
   \includegraphics[width=\textwidth]{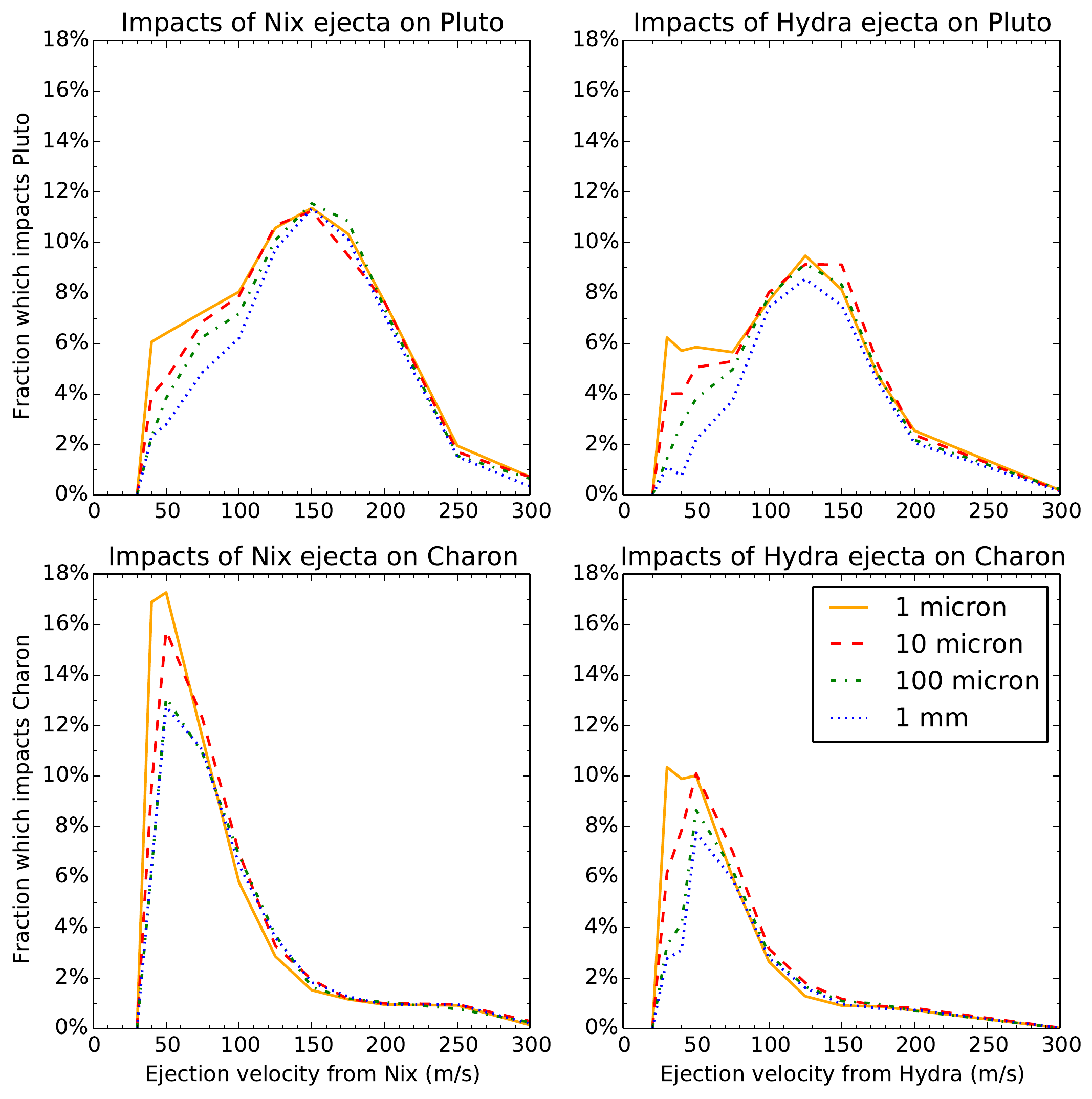}
   \caption[Fraction of ejecta which impacts without air drag]{ \label{fig:velocity_nodrag}
      Fraction of small satellite ejecta which impacts Pluto and Charon with no air drag as a function of ejection velocity and particle size.
   }
\end{figure}

For simulations without hydrodynamic drag from Pluto's atmosphere, Figure \ref{fig:velocity_nodrag} shows that
the main effect of radiation pressure was to 
increase the number of Pluto impacts by smaller dust particles.
These impacts were only produced by low velocity ejections (\textless100 m/s) which had a close encounter with Charon 
and/or Pluto, thus ejecting them into very eccentric orbits with apoapses close to Pluto's Hill sphere.
This allowed the larger particles to be perturbed by solar gravity into even more eccentric unbound orbits.
However, since the solar gravity and radiation pressures almost canceled out for the smaller particles,
they where able to make several of these eccentric orbits before eventually impacting Pluto or being scattered out of the system
(the eccentric orbits are not typically in the plane of Charon's orbit).
If only Pluto and Charon's gravity were included in the simulations, then all the dust particles on eccentric orbits
would behave similar to the 1 $\mu$m grains.
At higher ejection velocities than 100 m/s, close encounters would typically cause the particle to be immediately 
ejected from the system, and thus radiation pressure had little effect on those simulations.

\begin{figure}[p!]
   \includegraphics[width=\textwidth]{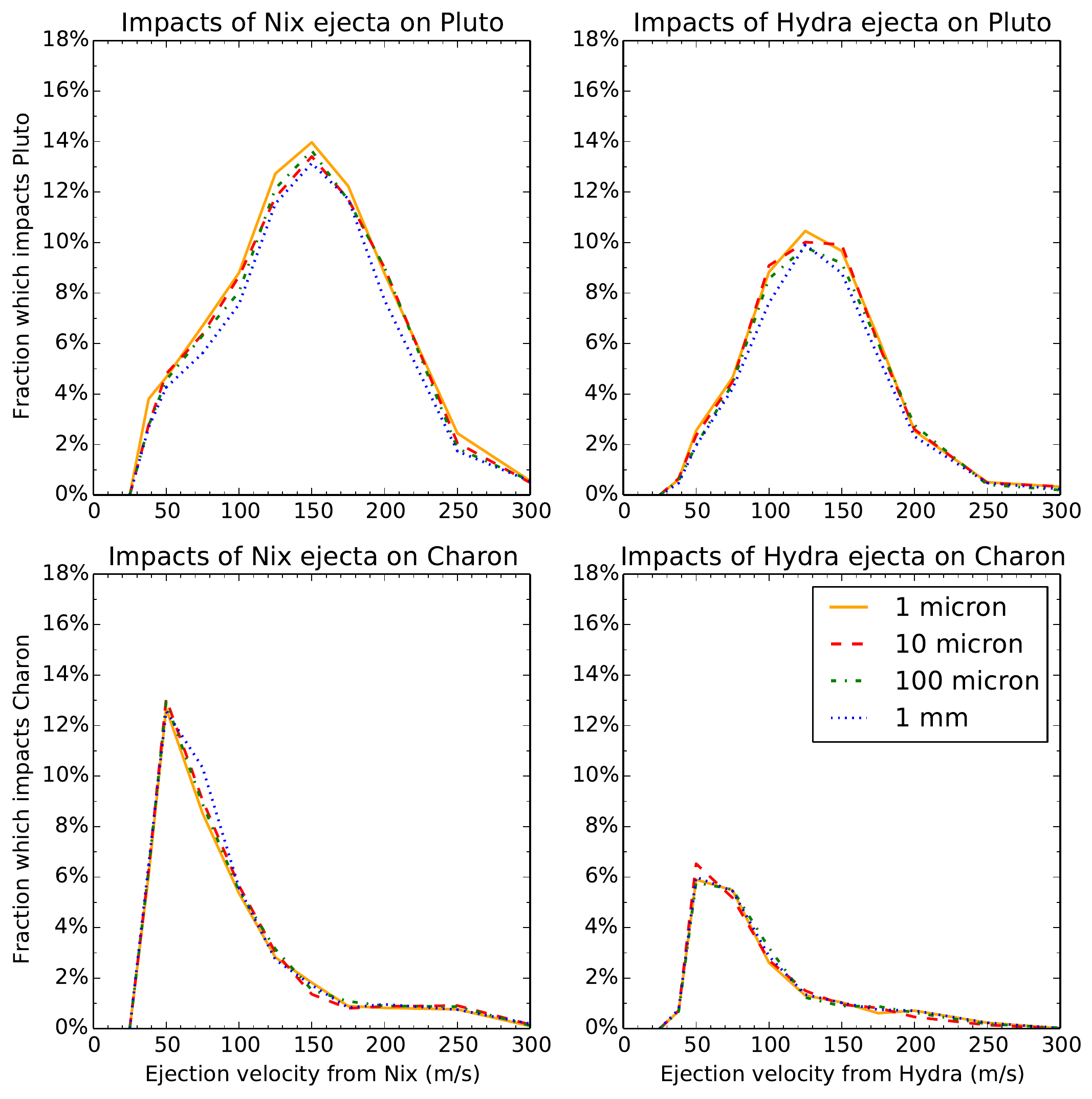}
   \caption[Fraction of ejecta which impacts with air drag]{ \label{fig:velocity}
      Fraction of small satellite ejecta which impacts Pluto and Charon with air drag from Pluto's atmosphere 
     as a function of ejection velocity and particle size.
   }
\end{figure}

\begin{figure}[p!]
   \includegraphics[width=\textwidth]{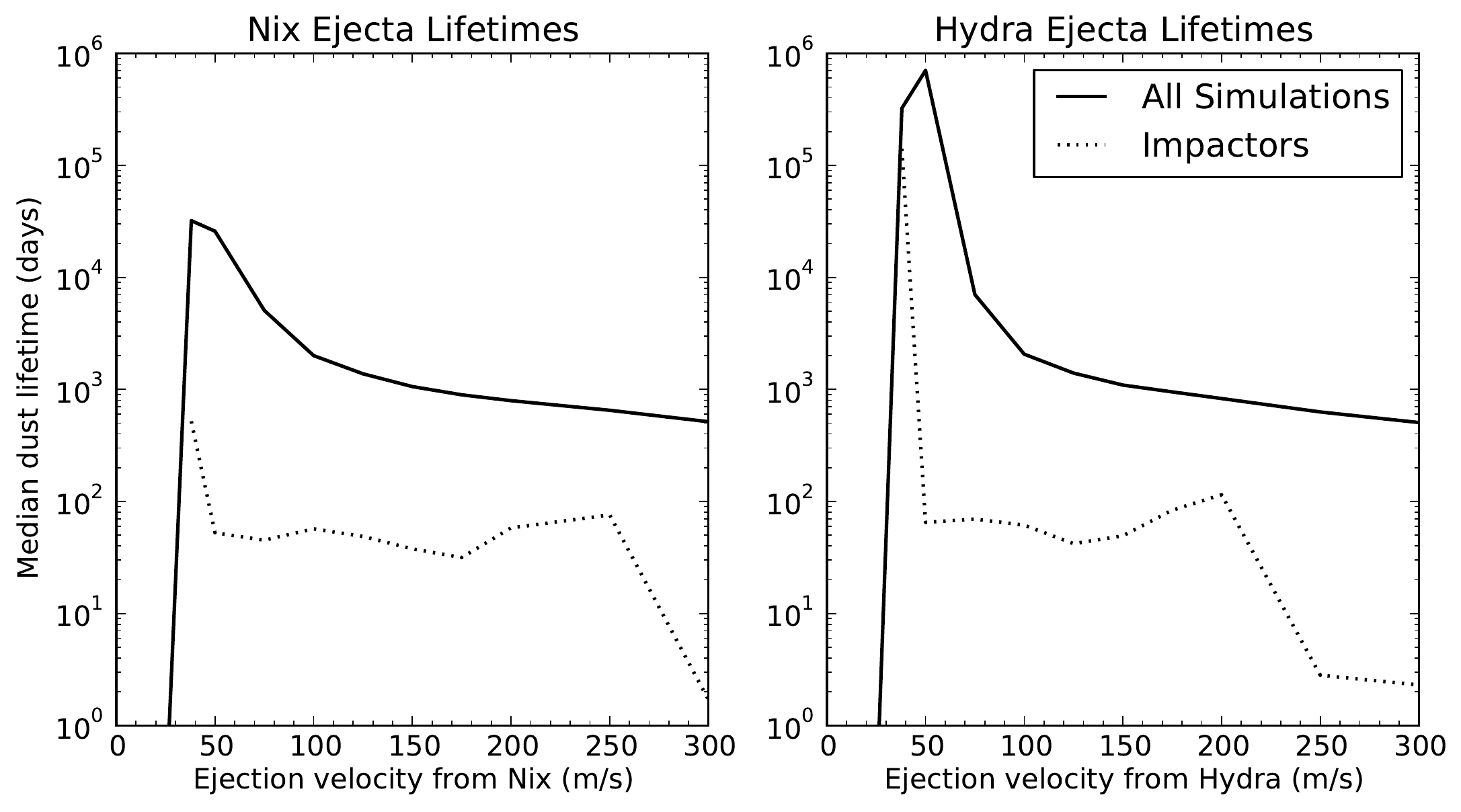}
   \caption[Lifetime of ejecta particles]{ \label{fig:lifetimes}
      Median lifetime of ejecta particles from Nix and Hydra before being ejected or impacting as a function of velocity.
      The solid line includes all simulations, while the dotted line includes only the simulations which resulted in
      the dust particle impacting either Pluto or Charon.
      Lifetimes for 10 $\mu$m ejecta are shown; other sizes are very similar.
   }
\end{figure}

Once drag is included in the simulations, Figure \ref{fig:velocity} shows that
these long-period eccentric trajectories are no longer possible (as they pass too close to Pluto),
and so radiation pressure and dust diameter again have little effect.

Figures \ref{fig:velocity_nodrag} and \ref{fig:velocity} show that in addition to reducing the amount of 
low-velocity ejecta which impacts both Pluto and Charon, Pluto's atmosphere also increases the amount of high-velocity ejecta 
which impacts Pluto.
These impactors are dust which flies close enough to Pluto where they experience some air drag which lowers the apoapse of their 
orbit around the Pluto system.
The dust particles frequently made several passes through the upper atmosphere until they lost enough energy to allow them to impact Pluto. 

\subsection{Impact Locations on Pluto and Charon}

\begin{figure}[p!]
   \includegraphics[width=\textwidth]{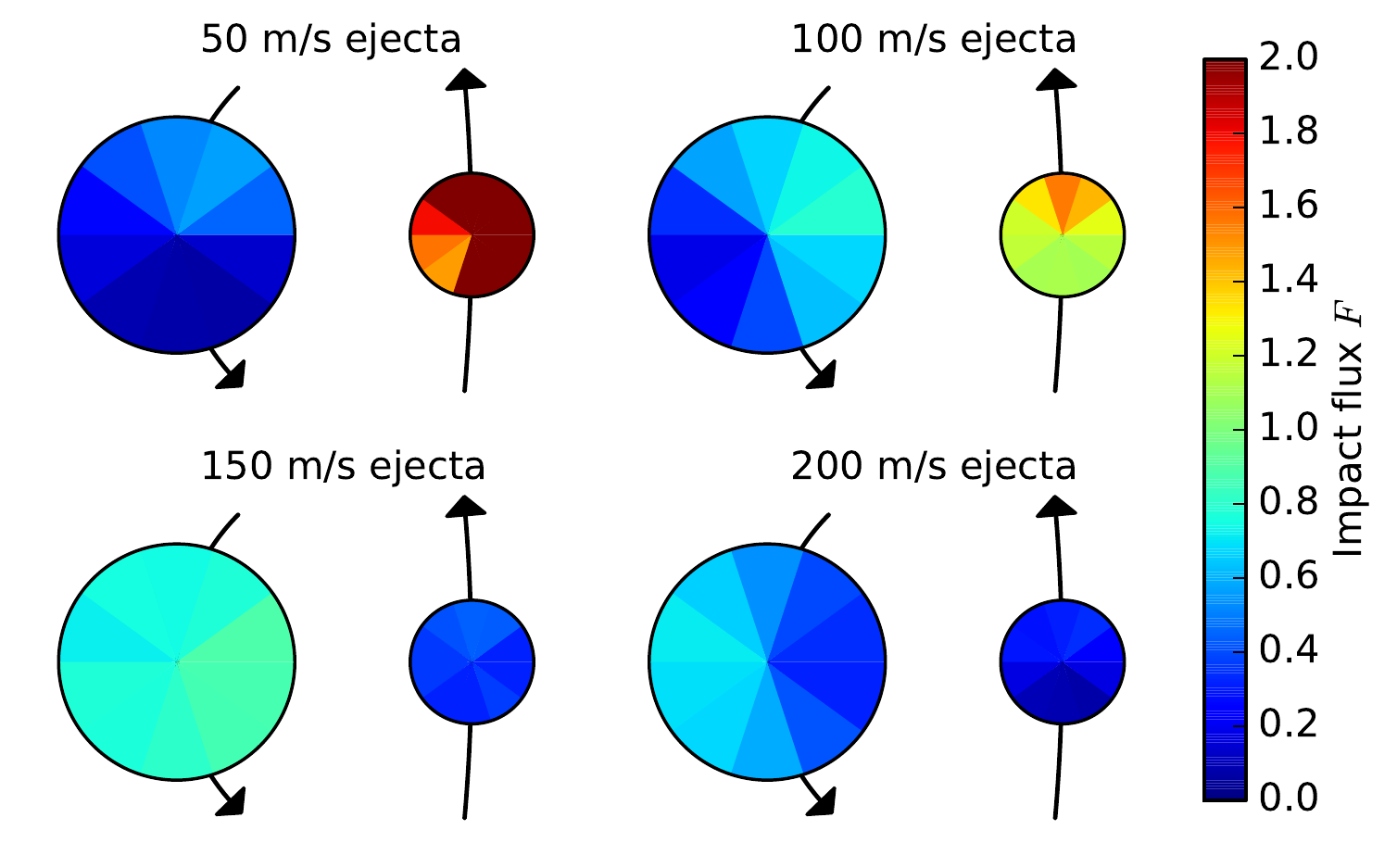}
   \caption[Longitude of dust impacts]{ \label{fig:pie1}
      Impact flux of 100 $\mu$m Nix ejecta on Pluto and Charon as a function of ejection velocity.
      The flux $F$ is the number of ejecta impacts per square kilometer per 10$^8$ ejecta particles.
      Pluto and Charon are to scale and viewed from their orbital north pole.
      A subsection of their orbits is shown to indicate the leading and trailing points on each body.
   }
\end{figure}

Figure \ref{fig:pie1} shows the longitudes of the impacts of 100 $\mu$m Nix ejecta on Pluto and Charon;
the other satellite ejecta simulations were similar.
Impacts on Charon are always primarily on the leading hemisphere due to Charon's own motion.
Low velocity ejecta also shows a preference for the anti-Pluto side, as dust ejected 
at these speeds can have an initial periapse exterior of Charon, but within Charon's sphere of influence,
causing them to be focused down onto Charon.

Impacts on Pluto center on two different points depending on their original ejection velocity.
The low velocity ejecta which impacts Pluto must be first scattered inward by Charon, 
which is most optimal for trajectories that impact Pluto's trailing point.
Because Pluto's orbit around the barycenter is very tight, the trailing point is only
about 75$^{\circ}$ east of the Charon-facing point.
The high velocity ejecta ($>$150 m/s), on the other hand, must impact Pluto directly 
without being scattered by Charon.
Thus, both the high velocity ejecta from the small satellites and the interplanetary dust particles
will primarily impact the anti-Charon side of Pluto.
In between these extremes, dust ejected at around 150 m/s can impact Pluto through 
a variety of trajectories, and thus produces a much more uniform distribution of impacts.

\subsection{Comparison of Impact Locations to Pluto's Albedo Features}

\begin{figure}[p!]
   \begin{center}
      \includegraphics[width=0.66\textwidth]{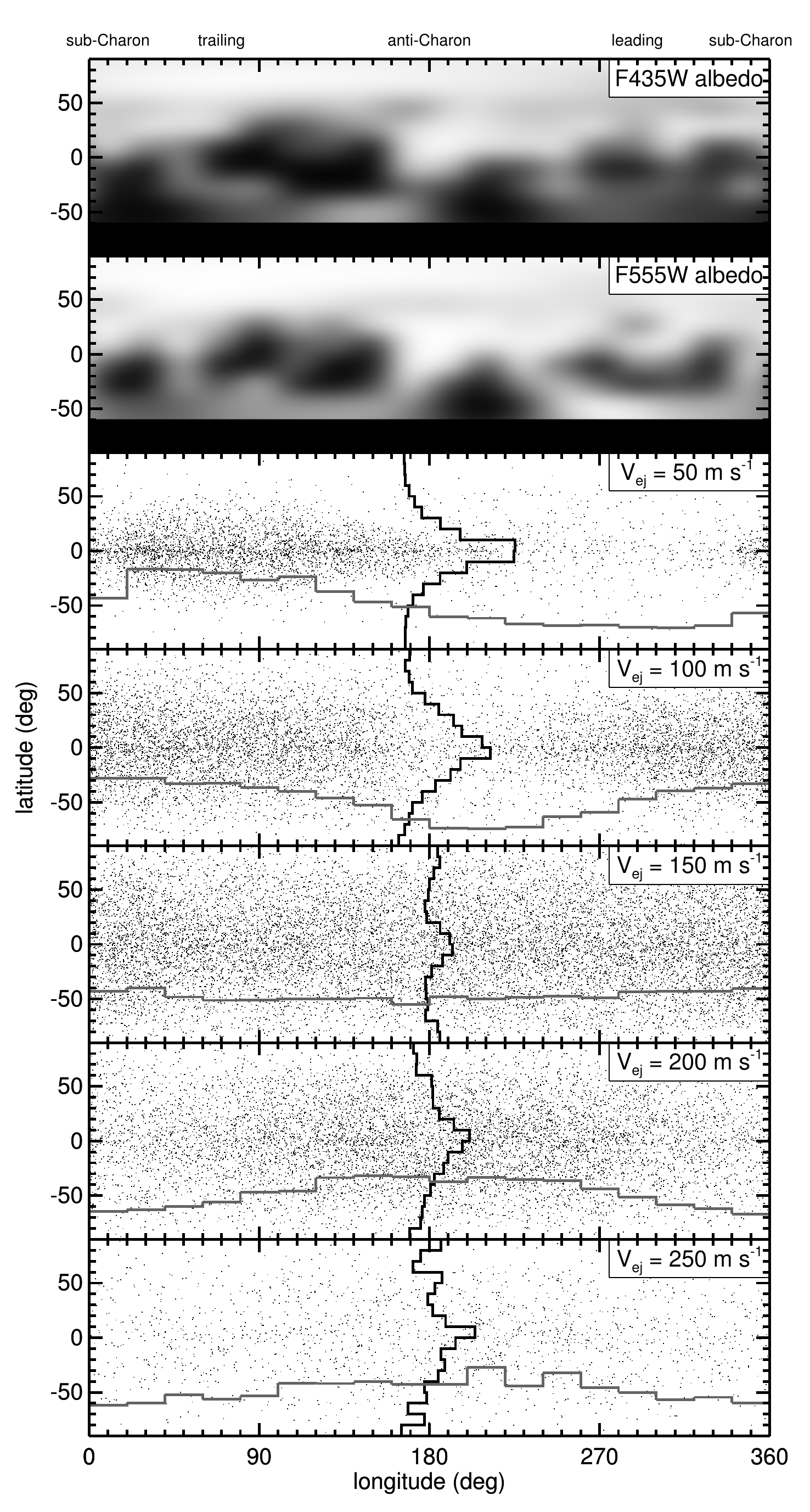}
   \end{center}
   \caption[Dust impact distribution on Pluto]{ \label{fig:maps}
      Comparison of \citet{Buie2010} albedo maps with spatial distribution of Nix ejecta impacts onto Pluto's surface.
      The progression from low to high velocity ejecta is shown from panels 3 to 7.  
      The histograms in each panel show the impact flux per unit area in 10$^{\circ}$ latitude 
      and 20$^{\circ}$ longitude bins (with arbitrary scaling).
   }
\end{figure}

Plotting the locations of ejecta impacts onto Pluto in map coordinates
reveals some interesting patterns, as shown in Figure \ref{fig:maps}.  Impacts by
lower velocity ejecta are concentrated toward the equator in addition
to their previously noted concentration on Pluto's trailing hemisphere.
Impacts from higher velocity ejecta are more uniformly distributed,
but still show higher fluxes at low latitudes.

An obvious question to ask is whether or not the distribution of
ejecta impacts onto Pluto correlates with the albedo of Pluto's
highly-variegated surface.  The most recent maps of Pluto were published
by \citet{Buie2010}, based on Hubble Space Telescope observations
obtained during 2002-2003.  That paper presents single scattering albedo
maps fitted to images taken in two filters, $F435W$ and $F555W$, with
central wavelengths of 435 and 555 nm, respectively.  At the time of
the observations, latitudes below about 60$^{\circ}$ south on Pluto were
unilluminated by the Sun and unobservable from Earth, so the maps only
cover territory north of that latitude (where north is defined by the
planet's angular momentum vector and east points in the direction of rotation).

To compare an albedo map with our calculated distribution of impacts,
we would ideally sample the impacts in exactly the same way as the albedo
map samples the surface, and then use the two sets of samples to compute
a correlation coefficient.  However, the way the \citet{Buie2010}
albedo maps were constructed is not obviously applicable to sampling our
impact distribution.  The \citet{Buie2010} map was based on 114 non-equal area
rectangular tiles drawn on a cylindrical projection of Pluto's surface.
The single scattering albedos of each tile were allowed to float as
114 free parameters in a Hapke model \citep{Hapke1993}.  To compute the
goodness of fit in each step of the iterative fitting cycle, the 114-tile
map was projected onto the surface of a sphere and then smoothed with a
two-dimensional Gaussian filter with a $FWHM$ of 15$^{\circ}$, before being
rotated to match the geometry of each HST image.  The solid angle of a
15$^{\circ}$ diameter circle on a sphere is about 0.21 steradians, implying
that only about 59 parameters would actually be needed to describe the
albedo distribution at that spatial resolution on a 4$\pi$ steradian
sphere, and only 44 would be needed to cover the area north of 60$^{\circ}$
south latitude.  So it appears to us that the \citet{Buie2010} maps
had too many free parameters.  If we sample our impact distribution with
more samples than justified by the spatial resolution of the albedo
map, potential correlations could appear more significant than they
really are, because we are effectively over-counting.  Additionally,
when projected to the sphere, many of the 114 tiles are highly distorted
with elongated shapes, especially near the pole.  Considering the Gaussian
smoothing, it makes more sense for us to use a grid composed of tiles
that are approximately equal-area and equant on the sphere, with sizes
consistent with the smoothing.  So we constructed a new grid by taking
an icosahedron and subdividing each of its 20 equilateral triangles
into 4 smaller equilateral triangles.  Normalizing the vertex vectors
to approximate a sphere slightly distorts these 80 triangular facets,
but they remain equant with near equal areas of about 0.16 steradians.
This solid angle matches that of a 13$^{\circ}$ diameter circle on the
sphere, similar to the width of the 15$^{\circ}$ Gaussian smoothing filter
used by \citet{Buie2010}.

\begin{figure}[p!]
   \includegraphics[width=\textwidth]{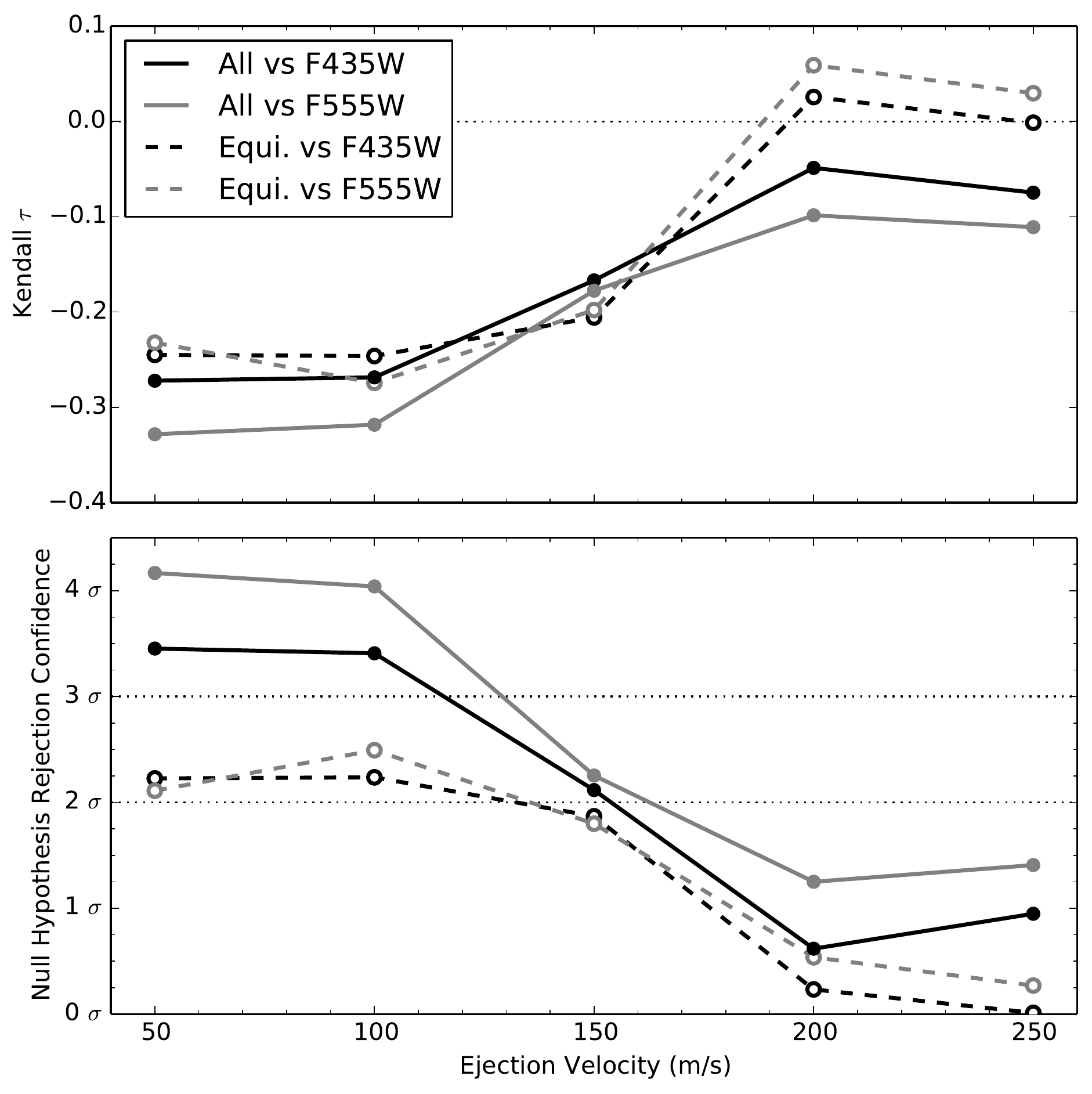}
   \caption[Correlation of Impacts to Albedo]{ \label{fig:sig}
      Upper panel: Kendall's $\tau$ rank correlation coefficient
      between Pluto's single scattering albedo maps at 435~nm (black curves)
      and 555~nm (grey curves) and ejecta departing Nix at five different
      velocities along the abscissa.  Solid lines
      represent a comparison of all latitudes north of 60$^{\circ}$ south, where
      the albedo maps cut off.  Dashed lines
      represent a comparison of only latitudes within 30$^{\circ}$ of the equator.
      Lower panel: The null hypothesis of no relation between albedo and impact
      flux can be rejected with greater confidence for low ejection velocities.
   }
\end{figure}

Since the distribution of impacts on Pluto appears symmetric across the
equator (see Figure \ref{fig:maps}), we doubled the impact counts by mirroring each
through the equator.  For each of the 80 triangular facets in our grid,
the number of impacts arriving in that facet was divided by the area of
the facet to obtain an impact flux per unit area.  We also computed the
mean single scattering albedo in each facet for the Gaussian-smoothed
$F435W$ and $F555W$ maps.  We could then compute correlation coefficients
between the impact fluxes and albedos, for all facets located north of
60$^{\circ}$ south latitude where the albedo maps are constrained.  Since
there is no reason to {\it a priori} expect a linear relationship between
single scattering albedo and impact flux, we used the nonparametric
Kendall $\tau$ rank correlation coefficient \citep{Press2007} to search
for potential relationships between these variables.  The results are
shown in Figure \ref{fig:sig}.  Impacts from higher velocity ejecta show $\tau$ values
near zero, as would be expected for unrelated variables, but the lower
velocity ejecta show a negative $\tau$, indicating an anti-correlation,
with lower albedos corresponding to areas receiving higher impact fluxes.
The null hypothesis of no relation between lower velocity ejecta impact
rates and both $F435W$ and $F555W$ single scattering albedos can be
rejected with more than three sigma confidence, as shown in the bottom
panel.  However, examining the distribution of impacts on Pluto, they
are strongly concentrated at low latitudes.  The dark albedo features
in the \citet{Buie2010} maps are mostly confined to low latitudes.
Pluto's high albedo north pole receives few low velocity ejecta impacts,
while its lower average albedo equatorial regions receive more.
To see how much of the albedo-impact anti-correlation comes from the
longitudinal distribution, we also evaluated Kendall's $\tau$ for just
the facets located within 30$^{\circ}$ of the equator.  Those $\tau$ values
are similarly negative for low velocity ejecta, as indicated by the 
curves in Figure \ref{fig:sig},
owing to Pluto's largest dark region coinciding with its trailing
hemisphere where the flux of low velocity ejecta is highest.  The null
hypothesis of no relation between impact flux and albedo is rejected
with less confidence using only the low-latitude regions, but it still
exceeds two sigma, suggesting that this is not just a latitude effect.

\section{Discussion}

All of the simulations in this study were scaled relative to the dust production rate from the small satellites.
This rate can in turn be broken down into the rate at which impacts occur on the small satellites and
the efficiency of the impacts in producing ejecta which is faster than the local escape velocity.

The size of the small satellites is not presently known.
\citet{Tholen2008} dynamically fit masses for Nix and Hydra and assumed they have Charon-like densities to derive equivalent spherical
diameters of 88 and 72 km, respectively, albeit with considerable uncertainty.
Accounting for shape effects and a lower density in line with observed objects of this size \citep{Mommert2012},
we assume a cross-sectional diameter of 100 km.
The impact flux on Pluto was estimated by \citet{deElia2010} as following a power law with approximately 3000 impacts of objects larger than 
1 km over the past 3.5 Gyr.
Scaling this flux to the cross-sectional areas of the small satellites, the largest plausible impactor is less than 2 km across.
The median encounter velocity of the Pluto system with other 3:2 resonant objects is 1.2 km/s, and these objects
probably dominate the impact flux \citep{Delloro2001}.
A 2 km object impacting a 55 km satellite at 1.2 km/s has an impact energy of approximately 3.5e5 erg/g.
From Figure 2 in \citet{Stewart2009}, this is not a disruptive impact, even assuming the satellite is a rubble pile.

\citet{Stewart2009} also provide an estimate of the relative mass of the largest fragment following an impact.
If we assume that the remaining material is all ejecta, and that half of that ejecta leaves the small satellite, then the amount of
dust ejected from the satellite per impact can be estimated.
Combining this with the scaled flux from \citet{deElia2010} and integrating over impactor sizes from 1 m to 2 km provides 
an average total dust production rate of about 2 g/s per small satellite.
This is lower than the production rate of 16 g/s estimated by \citet{Durda2000} for a 100 km Plutino, 
but reflective of the uncertainty in the calculation.
Much of this uncertainty is due to the fact that the dust is not produced at a constant rate.
Most of the dust is ejected by the largest few impacts on the small satellites (100 m to 1 km impactors).
Smaller impacts are more frequent, but produce much less dust overall.

If we assume that most of the dust is ejected at 1-10\% of the impact velocity \citep{Hartmann1985},
then 10-100 m/s would be the most common ejection velocity range.
From Figure \ref{fig:velocity}, over this range on average about 6\% of ejecta from the small satellites reaches Pluto,
and about 7\% reaches Charon.
Combining this with the production rate above of 2 g/s and the surface areas of the objects,
the approximate ejecta dust flux on Pluto from either Nix or Hydra is about is $1\times10^{-17}$ kg s$^{-1}$ m$^{-2}$,
and $3\times10^{-17}$ kg s$^{-1}$ m$^{-2}$ on Charon.
Styx and Kerberos are dimmer than Nix and Hydra (and therefore assumed to be smaller), but Styx is quite close to Charon,
and so we estimate that the total dust flux from the four satellites is approximately three times the amount from Hydra.
The total dust flux on Pluto is then $3\times10^{-17}$ kg s$^{-1}$ m$^{-2}$, and $1\times10^{-16}$ kg s$^{-1}$ m$^{-2}$ on Charon.
If the dust were uniformly distributed across their surfaces and had a mean density of 0.5 g/cm$^3$,
then over the past 3.5 Gyr, Pluto should have accumulated about 1 cm of dust and Charon 3 cm.
However, as the dust does not fall uniformly on their surfaces (especially Pluto), local dust depths could be much higher.

\subsection{Albedo Effects of Dust on Pluto}

As noted above, the impact locations of the low velocity small satellite ejecta appear to be 
correlated to the locations of the dark features around Pluto's equator.
These areas are compositionally distinct, with much weaker volatile ice (CH$_4$ and N$_2$) 
spectral features than the rest of the surface \citep[and references therein]{Grundy2013}.
Volatile transport models generally assume a preexisting dark terrain at the dark albedo features, 
which allows the volatiles to concentrate away from them \citep{Young2013}.
The ejecta dust might be the reason these terrains are darker, rather than any underlaying geology.
In particular, \citet{Bottke2013} showed that only a few centimeters of dust from the
Jovian irregular satellites could explain the dark material on the surface of Europa.
And \citet{Spencer2010} showed that thermal segregation on Iapetus enhances the 
contrast difference between the leading and trailing hemispheres. 
The longitudinal concentration of the low-velocity ejecta on the trailing hemisphere could be especially 
indicative of the dust's role in producing the dark areas.
However, the correspondence between albedo and the dust impact locations is far from perfect, and so if the small satellite dust
does play a role in changing the albedo of Pluto's surface, it is undoubtedly complicated by the surface topology and
more complex thermodynamic effects.

NASA's \textit{New Horizons} spacecraft will fly through the Pluto system in July 2015,
passing within 12,000 km of Pluto and 28,000 km of Charon \citep{Guo2008}.
This will allow global panchromatic imaging of both objects \citep[{\textless}0.7 km/px for Pluto and
{\textless}1.4 km/px for Charon;][]{Young2008} and should enable some of these complications to untangled.
In particular, if the bright albedo region is geologically distinct, it could appear to be 
``painted terrain'', as seen on Iapetus \citep{Tamayo2011}.

\subsection{Micrometeorite Annealing of Ice on Charon}

Charon has crystalline water ice on its surface \citep{Brown2000,Buie2000,Dumas2001}.
This was not expected, as Charon's surface is sufficiently cold that any water ice should not be warm enough to
anneal any radiation damage from solar and galactic cosmic rays.
\citet{Cook2007} suggested micrometeorite impacts may produce enough heat to anneal the ice, but rejected this
in favor of cryovolcanism as the IDP impact flux did not appear to be sufficient to anneal the ice.
\citet{Porter2010} revisited the micrometeorite impact model for a number of outer solar system surfaces
and put a lower limit for the IDPs that Pluto encounters of 1.6$\times10^{-17}$ kg s$^{-1}$ m$^{-2}$ in
order to compete with an assumed solar UV amorphisation rate of 40,000 years.
The small satellite ejecta impacts less energetically than the IDPs, resulting in an annealing rate
for the small satellites that is about ten times slower per mass flux than the IDPs.
The ejecta dust impact flux estimated above would then be sufficient for micrometeorite annealing to be 
competitive with solar UV amorphisation.
Observations by the Ralph/LEISA spectrometer on \textit{New Horizons} may be able to test this \citep{Young2008}, 
especially if the leading hemisphere (which receives more small satellite dust, 
but approximately equal amounts of IDPs) appears more crystalline than the trailing hemisphere.

\section{Conclusions}

Through dynamical simulations, we have shown that dust ejected from the small satellites of the Pluto system
can impact the surfaces of either Pluto or Charon.
Dust ejected at lower velocities (\textless150 m/s) will preferentially impact Charon, but will also impact the trailing hemisphere of Pluto.
Dust ejected at higher velocities (\textgreater150 m/s) is more likely to impact Pluto, especially on the 
anti-Charon hemisphere.
High velocity interplanetary dust particles (\textgreater1 km/s) behave the same as the high velocity ejecta.
Charon receives more of the small satellite dust overall than Pluto, and those impacts are primarily on the leading 
hemisphere.
The low-velocity small satellite ejecta impacts Pluto in locations that correspond well to the dark albedo features 
observed on the equatorial areas of Pluto, implying that the dust may help to make those areas of Pluto dark.
Observations by the \textit{New Horizons} spacecraft may therefore show these regions painted with 
dark small satellite dust.

\section*{Acknowledgement}

This project was supported in part by NSF Planetary Astronomy Grant AST-1109872 
and also benefited from constructive feedback by two anonymous reviewers.
We thank Frank Timmes for donating to this project the required CPU time on the ASU Saguaro supercomputer.
This project also benefited from free and open source software, most notably the \textit{clang} C++ compiler,
the \textit{matplotlib} graphics library, and the \textit{TeX Live} latex package.

\pagebreak
\bibliographystyle{elsarticle-harv}
\bibliography{globalrefs}

\end{document}